\documentclass{amsart}

\addtocounter{secnumdepth}{1}

\usepackage{epstopdf}
\usepackage[bookmarksnumbered,colorlinks,bookmarks,citecolor=blue,linkcolor=blue,breaklinks,linktocpage]{hyperref}
\usepackage[all]{hypcap}
\usepackage{subfigure}

\usepackage{amsfonts}
\usepackage{amsmath}
\usepackage{amsxtra}
\usepackage{amssymb}
\usepackage{amscd}
\usepackage{array}
\usepackage{comment}
\usepackage{diagrams}
\usepackage{epsfig}
\usepackage[all]{xy}
\usepackage{pifont}
\usepackage{array}
\usepackage{enumerate}
\usepackage{longtable}
\usepackage{float}
\usepackage{multicol}
\usepackage{tabularx}
\usepackage{mathrsfs}

\theoremstyle{plain}
\newtheorem{theorem}{Theorem}[section]

\theoremstyle{definition}
\newtheorem{definition}[theorem]{Definition}

\theoremstyle{remark}

\newcommand{\rubato}{$\mbox{RUBATO}^{\mbox{{\tiny \circledR}}}$}

\newcommand{\ZZ}{\Bbb{Z}}

\newcommand{\nin}{\not\in}

\begin{document}

\title{Exploring Exotic Counterpoint Compositions}
\author{Octavio Agust\'{i}n-Aquino}
\address{Universidad Tecnol\'{o}gica de la Mixteca, Mexico}
\email{octavioalberto@mixteco.utm.mx}
\author{Jeffery Liu}
\address{College of Liberal Arts, University of Minnesota}
\email{liu00614@umn.edu}
\author{Guerino Mazzola}
\address{School of Music, University of Minnesota}
\email{mazzola@umn.edu}

\begin{abstract}
In this paper, first musical compositions are presented, which are created using the mathematical counterpoint theory of Guerino Mazzola and his collaborators. These compositions also use the \rubato\ software's components for counterpoint constructions. The present work aims at opening new ``exotic'' directions of contrapuntal composition in non-Fuxian worlds. The authors would like to receive first impressions about these compositions, which are available as scores and audio files.
\end{abstract}


\maketitle

\section*{Introduction}\label{intro}
A mathematical model of classical counterpoint for first species as codified by \cite{fux} was first published in \cite{cptEEG} and later described in detail in \cite{tom,tom2}. The original model was extended to five ``exotic'' counterpoint worlds, where the Fux theory model can be restated {\em mutatis mutandis} for five new dichotomies of intervals in the 12-tempered chromatic tuning. The context of 12-tempered tuning was extended to ``microtonal'' environments, together with a sophisticated embedding of contrapuntal theories in successively refined tuning spaces. The present state of these theories, together with their implementation in the composition software \rubato\ was described in \cite{cpworlds}.

In this paper we want to present and discuss a number of contrapuntal compositions that emerge from those exotic worlds. This approach should inspire composers to extend contrapuntal creativity to new aesthetic frontiers. We believe that this exploration may help create compositions of contrapuntal quality beyond the classical tradition, a quality that is guaranteed by its strict logic, but which is set up in a different categorization of ``consonant'' versus ``dissonant'' interval categories.

Recall that, historically speaking, the traditional consonant and dissonant intervals and the corresponding compositions evolved in a very long experimental unfolding of compositions from the early stage around 900 A.D. to the mature compositions of Giovanni Pierluigi da Palestrina in the 16th century, which testified the convergence of a contrapuntal interval classification to the Fuxian catechism. We believe that our new exotic worlds of counterpoint should also be subjected to an experimental development to understand their extended creative potential. Observe that the interval categories around 900 A.D. were understood from the mathematical music theory of Pythagorean proportions. The experimental development that converged in Palestrina's interval categories was a {\em musical research} that remarkably generated a mathematically exquisite consonance/dissonance dichotomy. Our own mathematical model started from this Fuxian canon and developed a new set of contrapuntal worlds. The next movement in this trajectory is again similar to the experimental phase after 900 A.D., it should deepen the musical understanding of contrapuntal worlds and converge to some future Palestrina, who would comprehend the essence and form of the variety of our six contrapuntal worlds. This is the vector we propose with this paper.

The paper is organized as follows: We first give a summary of the present theory; we then discuss its implementation in the Java-based composition software \rubato. In a third section, we present and discuss a number of compositions, which are enabled by the software implementation. A fourth section should enable the reader to comment upon the given compositions, and to suggest further developments in a poll/inquiry.

\section{Summary of the Theory of the Six Counterpoint Worlds}\label{CPT}
The present counterpoint model starts with the concept of an interval number dichotomy:

\begin{definition}\label{def:dichotomy}
A dichotomy (of interval numbers) in $\ZZ_{2n}$ is an ordered pair $(X/Y)$ of complementary subsets $X,Y\subset \ZZ_{2n}$ of equal cardinality $n$, i.e., $Y=\ZZ_{2n}-X$ and $card(X)=card(Y)=n$.
\end{definition}

The group $\ZZ_{2n}$ is interpreted as the space of interval numbers in the pitch class group $\ZZ_{2n}$. An interval is conceived as an element $z=x+\varepsilon.y\in\ZZ_{2n}[\varepsilon]$, the group of dual numbers over $\ZZ_{2n}$, where $x$ is the interval's basis, the cantus firmus note, whereas $y$ is its interval number. For $2n=12$, for example, the interval $4+\varepsilon 7$ would represent the fifth interval (interval number 7) starting at $4$ (pitch class of e).

Interval number dichotomies can be transformed under the affine automorphism group $\overrightarrow{GL}(\ZZ_{2n})$ of $\ZZ_{2n}$. If $g=T^t.u\in \overrightarrow{GL}(\ZZ_{2n})$, with translation $t$ and invertible multiplicator $u$, we set $g(X/Y)=(g(X)/g(Y))$. 

\begin{definition}\label{def:strong}
An interval dichotomy $(X/Y)$ is said to be {\em rigid} iff its fix point group is trivial, i.e., $g(X/Y)=(X/Y)$ iff $g=Id$. It is said to be strong iff it is rigid, and if there is an {\em autocomplementarity} $p\in \overrightarrow{GL}(\ZZ_{2n})$ such that $p(X/Y)=(Y/X)$. For strong dichotomies, their autocomplemetarity is uniquely determined.
\end{definition} 

Evidently, being strong for $(X/Y)$ is inherited to all of its orbit $\overrightarrow{GL}(\ZZ_{2n}).(X/Y)$, it is a property of the affine orbit of the strong dichotomy. For $\ZZ_{12}$, there are six classes of strong dichotomies. Here is their list, together with the number in the general classification of subsets of $\ZZ_{12}$ and their polarities:

\vspace{0.2cm}
\begin{table}
\begin{tabular}{|c|c|c|c|}
\hline
Number & Dichotomy & Forbidden parallels & polarity\\
\hline
64&$(I/J)=(\{2,4,5,7,9,11\}/\{0,1,3,6,8,10\})$ & $5$, $11$ &$T^5.11$\\
68&$(\{0,1,2,3,5,8\}/\{4,6,7,9,10,11\})$ & $0$, $2$, $8$ &$T^6.5$ \\
71&$(\{0,1,2,3,6,7\}/\{4,5,8,9,10,11\})$ & None & $T^{11}.11$\\
75&$(\{0,1,2,4,5,8\}/\{3,6,7,9,10,11\})$ & None & $T^{11}.11$\\
78&$(\{0,1,2,4,6,10\}/\{3,5,7,8,9,11\})$ & None & $T^9.11$\\
82&$(K/D)=(\{0,3,4,7,8,9\}/\{1,2,5,6,10,11\})$ & $7$ &$T^2.5$\\
\hline
\end{tabular}
\caption{The 6 Dichotomy Classes of $\ZZ_{12}$}
\label{tab:dichotomyClasses}
\end{table}
\vspace{0.2cm}

The column ``forbidden parallels'' will be explained below. It is remarkable that
\begin{enumerate}
\item the last strong dichotomy 82 is the classical Fux dichotomy with the six consonant interval numbers (multiples of semitones) $0,3,4,7,8,9$, whereas
\item the first dichotomy 64 is defined by the six proper intervals in the ionian scale from the tonic.
\item Dichotomy 78 is generated by the six pitches of Scriabin's mystic chord.
\end{enumerate}

For the other three strong dichotomies we actually have no musical rationale.

Both dichotomy 64 and 82 are geometrically in distinguished positions if we draw them as partitions of the third torus $\ZZ_3\times \ZZ_4$ which is the Sylow representation of $\ZZ_{12}$. More precisely, define the {\em diameter} of a strong dichotomy $(X/Y)$ with polarity $p$ by $\delta(X/Y) =\frac{1}{2}\sum_{u,v\in X}d(u,v)$, and the {\em span} by $\sigma(X/Y)=\sum_{u\in X}d(u,p(u))$. Here $d(u,v)$ is the third distance of two pitch classes in the third torus, i.e., the minimal number of thirds connecting $u$ and $v$. With these distance numbers it turns out that dichotomy 82 has minimal diameter and maximal span, whereas dichotomy 64 has maximal diameter and minimal span, they are in a polar position as strong dichotomies on the geometry of the third torus. In particular, the Fux dichotomy 82 is an optimal separator of its two defining sets of consonances versus dissonances. This geometric poperty exhibits the Fux dichotomy from all strong dichotomies. In particular, the fourth, given by $5$ semitones, is a dissonance and not a consonance as would be defined in the acoustical criteria as defined by the Pythagorean theory.

The next step of our theory is designed to exhibit the allowed successors of consonant intervals $x+\varepsilon.k$ in first species counterpoint. To this end, we denote by $X[\varepsilon]$ the set of consonant intervals with interval numbers i $X$. We now apply affine automorphisms $g$ of $\ZZ_{2n}[\varepsilon]$ to the interval dichotomies $X[\varepsilon]=\ZZ_{2n}+\varepsilon.X$. 

\begin{definition}
Let $\Delta[\varepsilon]$ be the dichotomy $(X[\varepsilon]/Y[\varepsilon])$ deduced from the strong dichotomy $(X/Y)$ with polarity $p=T^u.v$. Fix a consonant interval $\xi = x+\varepsilon.k, k\in X$. Then an automorphism $g$ is {\em contrapuntal for $\xi$} if
\begin{enumerate}
\item 
$\xi\nin g(X[\varepsilon]$,
\item
$p^x_{\Delta}$ is a polarity of $g.\Delta[\varepsilon]$,
\item
The cardinality of $g(x[\varepsilon])\cap X[\varepsilon])$ is maximal among those $g$ with the properties (1)and (2).
\end{enumerate}
\end{definition}
Here, we define $p^x_{\Delta}=T^{x(1-v)+\varepsilon.u}.v$. This polarity maps the ``tangent space'' $y+\varepsilon.\ZZ_{2n}$ to the tangent space $x+v(y-x)+\varepsilon.\ZZ_{2n}$.

This definition means that $g$ deforms the given dichotomy $\Delta[\varepsilon]$ such that the starting interval $\xi$ is a deformed dissonance, and we are looking for all deformed consonances. The idea is to associate this deformed dissonance $\xi$ with a maximal set of deformed consonances.

\begin{definition}\label{def:successor}
 For a strong dichotomy $\Delta[\varepsilon]$ and a consonant interval $\xi\in X[\varepsilon]$, an interval $\eta$ is called an admitted successor of $\xi$ if it is contained in the intersection $g(X[\varepsilon])\cap X[\varepsilon]$ for a contrapuntal automorphism $g$ for $\xi$.
\end{definition}

This definition enables us to calculate all possible successors of consonant intervals for a given strong dichotomy. The explicit calculation of successors is given by an algorithm, which was described by Jens Hichert \cite{tom}.

\begin{theorem}[Gro{\ss}er Kontrapunktsatz] Let $\Delta = (X/Y)$ a strong interval dichotomy in $\mathbb{Z}_{12}$, and 
let $\xi$ be a consonant interval. Then $\xi$ has at least $42$ admissible successors, and it has at least one admissible
successor even if we choose the cantus firmus of the successor beforehand.
\end{theorem}

The forbidden parallel sucessors are listed in the above table. In particular, one recognizes that for the Fux dichotomy 82, parallels of fifths are the only forbidden cases, and this confirms these cases as prescribed by the classical Fux system. 

{\em Summarizing, the dissonant fourth and the forbidden parallels of fifths are a consequence of our mathematical model for the Fux dichotomy in $\ZZ_{12}$.}

As implied in the preceding discussion, the theory is generalizable to any equally tempered $2n$-scale, with $2n>4$. We have the following result (see \cite{cpworlds}).

\begin{theorem}[Kleiner Kontrapunktsatz]\label{T:KKS} Let $\Delta = (X/Y)$ a strong interval dichotomy in $\mathbb{Z}_{2n}$, $\xi$ be a consonant interval and $N$ the number of admitted successors of $\xi$ determined by a specific contrapuntal automorphism. The following inequality holds
\[
 n^{2}\leq N \leq 2n^{2}-n.
\]
\end{theorem}

\begin{definition}
Let $\Delta_{1} = (X_{1}/Y_{1})$ and $\Delta_{2} = (X_{2}/Y_{2})$ two strong dichotomies in ambient spaces $\mathbb{Z}_{2n_{1}}$ and $\mathbb{Z}_{2n_{2}}$, respectively. A \emph{morphism} between these dichotomies is an affine morphism of modules $\phi:
\mathbb{Z}_{2n_{1}}\to \mathbb{Z}_{2n_{2}}$ such that $\phi(X_{1})\subseteq X_{2}$, and the square
\[
	\begin{CD}
	\mathbb{Z}_{2n_{1}}@>\phi>>\mathbb{Z}_{2n_{2}}\\
	@Vp_{\Delta_{1}}VV @VVp_{\Delta_{2}}V\\
	\mathbb{Z}_{2n_{1}}@>\phi>>\mathbb{Z}_{2n_{2}}\\
	\end{CD}
\]
commutes.
\end{definition}

A special case occurs when $n_{2} = 2n_{1}$ and $\phi$ is the multiplication by $2$, in which case we simply write
\[
 \Delta_{2n} \rightarrowtail \Delta_{4n}
\]
for the morphism.

\begin{theorem}\label{T:Embedding}
There exists an infinite sequence of strong dichotomies $\{\Delta_{2^{n}\cdot 3}\}_{n=1}^{\infty}$ (with
$\Delta_{2^{n}\cdot 3}$ a subset of $\mathbb{Z}_{2^{n}\cdot 3}$) such that
\[
 \Delta_{6} \rightarrowtail \Delta_{12} \rightarrowtail \Delta_{24} \rightarrowtail \cdots
\]
with polarities
\[
 p_{n} = T^{2^{n-1}}(4^{\lceil\frac{n}{2}\rceil}+1).
\]
\end{theorem}

It is to be noted that Theorem \ref{T:Embedding} allows us to relate the consonances within a scale with the set of consonances in another scale but with the double of tones, and such that the former are preserved within the latter, yet it does not explicitly reveal anything about the relationship between the corresponding first-species counterpoints. There are some indications of how to do this in \cite{Vallarta}, but in general is an open topic. Another direction towards the relationship between counterpoint with different sets of consonances stems from the notion of \emph{counterpoint world} that we proceed now to explain.


\begin{definition}
Let $\Delta = (X/Y)$ be a strong dichotomy of $\mathbb{Z}_{2k}$ with polarity function $p_{\Delta}$ with $2k\geq 6$ and
$S\subseteq \mathbb{Z}_{2k}[\epsilon]$. The \emph{global polarity function} $p_{\Delta}^{\bullet}:\mathbb{Z}_{2k}[\epsilon]
\to \mathbb{Z}_{2k}[\epsilon]$ is defined by
\[
 p_{\Delta}^{\bullet}(x+\epsilon.i) = x+\epsilon.p_{\Delta}(i),
\]
and the \emph{closure}
\[
 \overline{S} := S\cup p_{\Delta}^{\bullet}(S).
\]

A \emph{counterpoint world} $\mathcal{W}$ is a triple $(\kappa,\sigma,p_{\Delta}^{\bullet})$, where $\kappa$ and
$\sigma$ are the characteristic functions
\begin{align*}
 \kappa:\overline{S}&\to \{0,1\},\\
 \xi &\mapsto\begin{cases}
 1, & \xi\in X[\epsilon]\cap \overline{S},\\
 0, & \text{otherwise},
 \end{cases}
\end{align*}
and
\begin{align*}
\sigma:\overline{S}\times\overline{S}&\to\{0,1\},\\
(\xi_{0},\xi_{1})&\mapsto\begin{cases}
 1, & \xi_{1}\text{ is an admitted succesor of }\xi_{0},\\
 0, & \text{otherwise}.
\end{cases}
\end{align*}
\end{definition}

Thus a counterpoint world is a set of consonances with its respective dissonances (defined by a polarity)
where you know when a consonance is a valid successor of another.

\begin{definition}
Given the mathematical worlds $\mathcal{W}_{1}=(\kappa_{1},\sigma_{1},p_{\Delta_{1}}^{\bullet})$ and
$\mathcal{W}_{2}=(\kappa_{2},\sigma_{2},p_{\Delta_{2}}^{\bullet})$ with respect to the sets
$\overline{S}_{1}=\mathrm{dom}(\kappa_{1})$ and $\overline{S}_{2}=\mathrm{dom}(\kappa_{2})$.
A \emph{counterpoint world morphism} is a function $\psi:\overline{S}_{1}\to\overline{S}_{2}$ that
is compatible with the consonances and dissonances, the admitted successors and the polarity, i.e.,
the following diagrams commute
\[
\xymatrix{
\overline{S}_{1}\ar[rr]^{\psi}\ar[dr]_{\kappa_{1}} & & \overline{S}_{2} \ar[dl]^{\kappa_{2}}\\
&\{0,1\}
},
\]
\[
\xymatrix{
\overline{S}_{1}\times \overline{S}_{1}\ar[rr]^{\psi\times\psi}\ar[dr]_{\sigma_{1}} & & \overline{S}_{2}\times \overline{S}_{2} \ar[dl]^{\sigma_{2}}\\
&\{0,1\}
},
\]
\[
\xymatrix{
\overline{S}_{1}\ar[r]^{\psi}\ar@{<->}[d]_{p^{\bullet}_{\Delta_{1}}} & \overline{S}_{2}\ar@{<->}[d]^{p^{\bullet}_{\Delta_{2}}}\\
\overline{S}_{1}\ar[r]^{\psi} & \overline{S}_{2}
}.
\]
\end{definition}

A counterpoint world morphism is, in essence, a coherent way to relate consonant and dissonant
intervals from one world into another (it could be from a world into itself, and not necessarily in
a trivial way). The counterpoint worlds endowed with these morphisms define a category.

Within a counterpoint world, as Giovanni Battista Martini noted in the eighteenth century, it
 is easier to pay attention to the forbidden steps instead of the allowed ones. 
 
\begin{definition}
 Let $\mathcal{W} = (\kappa,\sigma,p_{\Delta}^{\bullet})$ be a counterpoint world. Its associated
 \emph{strict digraph} is the directed graph $D$ with vertices
 \[
  V(D):=\kappa^{-1}(1)
 \]
 and the arrows that signal the forbidden steps
 \[
  A(D):=\sigma^{-1}(0)\cap(\kappa^{-1}(1)\times\kappa^{-1}(1)).
 \]
\end{definition}

For two strict digraphs $D_{1}$ and $D_{2}$, a strict morphism is a map $\phi:V(D_{1})\to V(D_{2})$ such that
\[
 (\xi_{0},\xi_{1})\in A(D_{1}) \iff (\phi(\xi_{0}),\phi(\xi_{1}))\in A(D_{1}).
\]

It can be proved that a strict morphisms induce counterpoint world morphisms (see \cite[Chapter 4]{cpworlds}) in
a unique way, and actually the former are used to streamline the computation of the later, and it is implemented
in the BollyMorph rubette.

\section{Summary of the Theory's Java Implementation on the Rubato Software}\label{IMPL}

\subsection{Computer-assisted Counterpoint with the \rubato \ Software.}
Experimentation with the counterpoint theory in the \rubato \ software requires the installation of the BollyFux plug-in. 
Compositional and computational tasks can then be performed by adding components called \textit{rubettes}, selecting their properties, and establishing links to allow transfer of information between their inputs and outputs. A complete documentation and installation instructions for the BollyFux rubettes is given in Chapter 6 of \cite{cpworlds}. A working example for computer-assisted composition in the first species style with playback and output to a MIDI file will be outlined below.
\subsubsection{Forms.}
Three types of data are exchanged between the BollyFux rubettes: \texttt{Score}, \texttt{StrongDichotomy}, and \texttt{Counterpoint}.\\
\\ 
The \texttt{Score} form represents a musical score, containing standard MIDI information such as pitch, duration, onset, loudness, and voice. Several built-in rubettes under the Score group allow for the manipulation of \texttt{Score} forms. The MidiFileIn rubette reads a MIDI file and converts it to a \texttt{Score} form, while the MidiFileOut rubette writes a score to a MIDI file. The Voice Splitter rubette splits a score in multiple voices into scores of individual voices, and the Voice Merger rubette combines up to 16 individual scores in a single voice into a single score. The ScorePlay rubette offers a built-in playback function with piano roll.\\
\\
The \texttt{StrongDichotomy} form represents a strong dichotomy of intervals as defined in Definition \ref{def:strong}, which defines the consonant contrapuntal intervals and their admitted successors. It is the output of the BollyWorld rubette, and is required in at least one of the inputs of the BollyComposer, BollyCarlo, AnaBollyser, BollyMorpher, and Counterpointiser rubettes.\\
\\
The \texttt{Counterpoint} form represents a succession of counterpoint intervals in $\ZZ_{12}[\epsilon]$. Note that this form only encodes information about pitch and interval up to octave equivalence, and not rhythm, onset, loudness, or absolute pitch. This format is more suitable for computation and transformation. The Counterpointiser and DeCounterpointiser rubettes convert between \texttt{Counterpoint} form and two \texttt{Score} forms representing the cantus firmus and discantus. It is one of the outputs of the BollyComposer, BollyCarlo, BollyMorpher, Counterpointiser rubettes, and it is required in at least one of the inputs of the 
AnaBollyser, BollyMorpher, and DeCounterpointiser rubettes.

\subsubsection{Network.}

\begin{figure}\label{fig:rubato_example}
\centering
\includegraphics[scale=0.4]{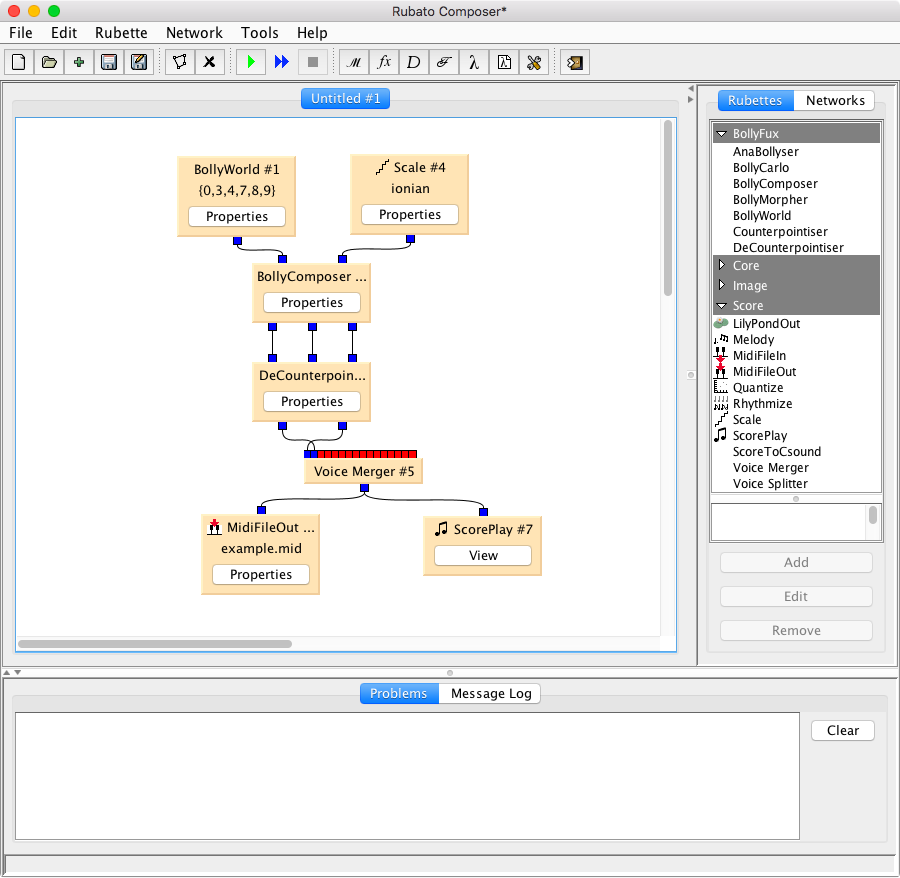}
\caption{A \rubato\ network for computer-assisted counterpoint}
\end{figure}

We give a complete recipe to build a network for computer-assisted composition in the first species style with playback and output to a MIDI file, see Figure \ref{fig:rubato_example}. 
Add the following BollyFux rubettes:
\begin{itemize}
\item BollyWorld
\item BollyComposer
\item DeCounterpointiser
\end{itemize}
and add the Score rubettes:
\begin{itemize}
\item Scale (optional)
\item Voice Merger
\item MidiFileOut
\item ScorePlay
\end{itemize}
Make the following links. Inputs and outputs are numbered from left to right starting from 0.
\begin{itemize}
\item BollyWorld output 0 $\rightarrow$ BollyComposer input 0
\item Scale output 0 $\rightarrow$ BollyComposer input 1 (optional)
\item BollyComposer output 0 $\rightarrow$ DeCounterpointiser input 0
\item BollyComposer output 1 $\rightarrow$ DeCounterpointiser input 1
\item BollyComposer output 2 $\rightarrow$ DeCounterpointiser input 2
\item DeCounterpointiser output 0 $\rightarrow$ Voice Merger input 1
\item DeCounterpointiser output 1 $\rightarrow$ Voice Merger input 0
\item Voice Merger output 0 $\rightarrow$ MidiFileOut input 0
\item Voice Merger output 0 $\rightarrow$ ScorePlay input 0
\end{itemize}

The properties of some rubettes will need to be set, as outlined below. 

\subsection{BollyWorld: Selection of a Strong Dichotomy.}
The BollyWorld rubette defines a strong dichotomy which is selected by the user in its properties. The default option, ``Topos of Music'', allows the user to select one of six representatives for each strong dichotomy class of $\ZZ_{12}$ as given by its number in Table \ref{tab:dichotomyClasses}. The ``Custom'' option allows the user to search through every member of every strong dichotomy class of any even octave division $\ZZ_{2n}$. Note that only $\ZZ_{12}$ will work with the BollyComposer rubette. The output of the rubette is a \texttt{StrongDichotomy} form.

\subsubsection{Scale: Restriction of Pitches.}
The built-in Scale rubette allows the user set a scale to which the composition will be restricted. In the properties of the scale, the user may select the number of notes, the root note in MIDI pitch, and the interval in semitones between successive scale degrees. The diatonic modes are preset options. The output of the rubette is a \texttt{Score} form representing the scale across the entire MIDI range.

\subsubsection{BollyComposer: Composition of a First Species Counterpoint.}
The BollyComposer rubette allows the user to make a simple composition with counterpoint rules in the first species style in two voices. It requires a strong dichotomy as its first input in order to define the consonant intervals and permitted progressions. Optionally, a scale may be connected as a second input, which restricts the notes of both cantus firmus and discantus. Note that once a link is established from the BollyWorld or Scale rubette, the network must be run once to apply or update the composition rules or scale restrictions. Once this is accomplished, the user may specify the length of the composition and create a composition by dragging each note up or down to different pitches on the staves for the two voices. The status box will alert the user whether the interval between the cantus firmus and discantus note is consonant or dissonant, and if it is consonant, whether it is a admitted successor according to Definition \ref{def:successor}. The three outputs of the BollyComposer rubette are a \texttt{Counterpoint} form representing the composition, and two \texttt{Score} forms for the cantus firmus and discantus voice.

\subsubsection{ScorePlay and MidiFileOut: Playback and Output.}
In the properties of MidiFileOut, select a MIDI file for the two voice to be written. After a composition is created in BollyComposer and the network is run once, the output will appear in the selected MIDI file and be available for playback in ScorePlay.

\section{Examples of Compositions for Exotic Counterpoint Worlds}\label{EXPL}

\subsection{Examples by Octavio Agustín-Aquino}\label{EXPL_OCT}

\begin{figure}[h!]\label{fig:volterra}
\parbox{2.8in}{
\includegraphics[page=1,scale=0.37]{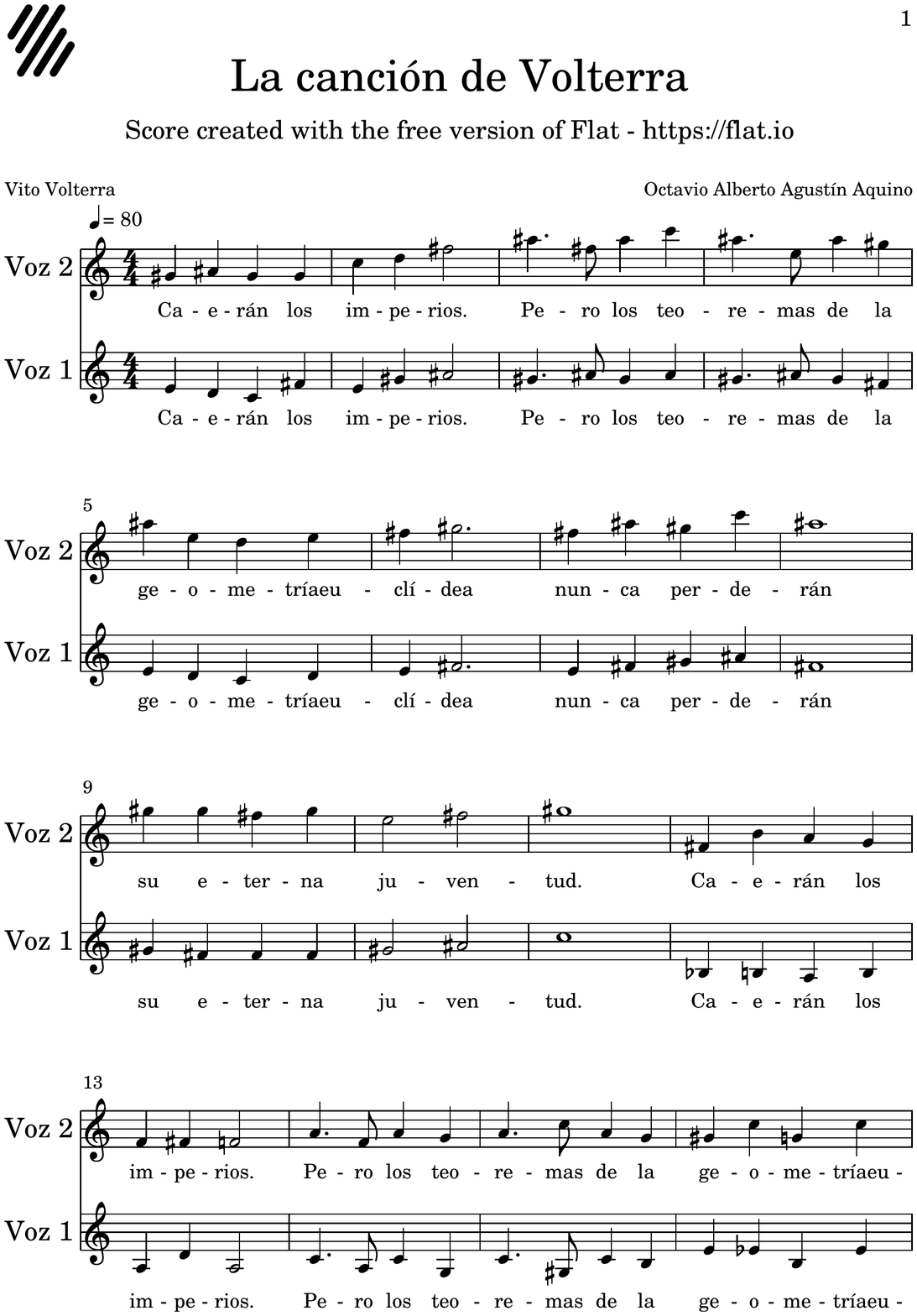}
}
\parbox{2.8in}{
\includegraphics[page=2,scale=0.37]{volterra.pdf}
}
\caption{La canci\'{o}n de Volterra, op. 38, for two voices. Copyright by Octavio Agust\'{i}n-Aquino}
\end{figure}

An initial first author's true attempt to write a piece of music\footnote{Audio files for all the following compositions are available at https://sites.google.com/site/elvotobatracio/musica.} in a counterpoint world different from the Fuxian one was \textit{La canci\'{o}n de Volterra}, op. 38 (Figure \ref{fig:volterra}). It is written for two voices, using as lyrics a free translation of a quote ascribed to Vito Volterra: ``Caer\'{a}n los imperios, pero los teoremas de la geometría eucl\'{i}dea nunca perder\'{a}n su eterna juventud''. The initial counterpoint was transformed using the BollyFux rubettes by Julien Junod from the $(K/D)$ world into the world $(\{0,2,4,6,8,11\}/\{1,3,5,7,9,10\})$, which is within the orbit of the mystical world.

\begin{figure}[h!]\label{fig:kontrapunktsatz}
\parbox{2.8in}{
\includegraphics[page=1,scale=0.37]{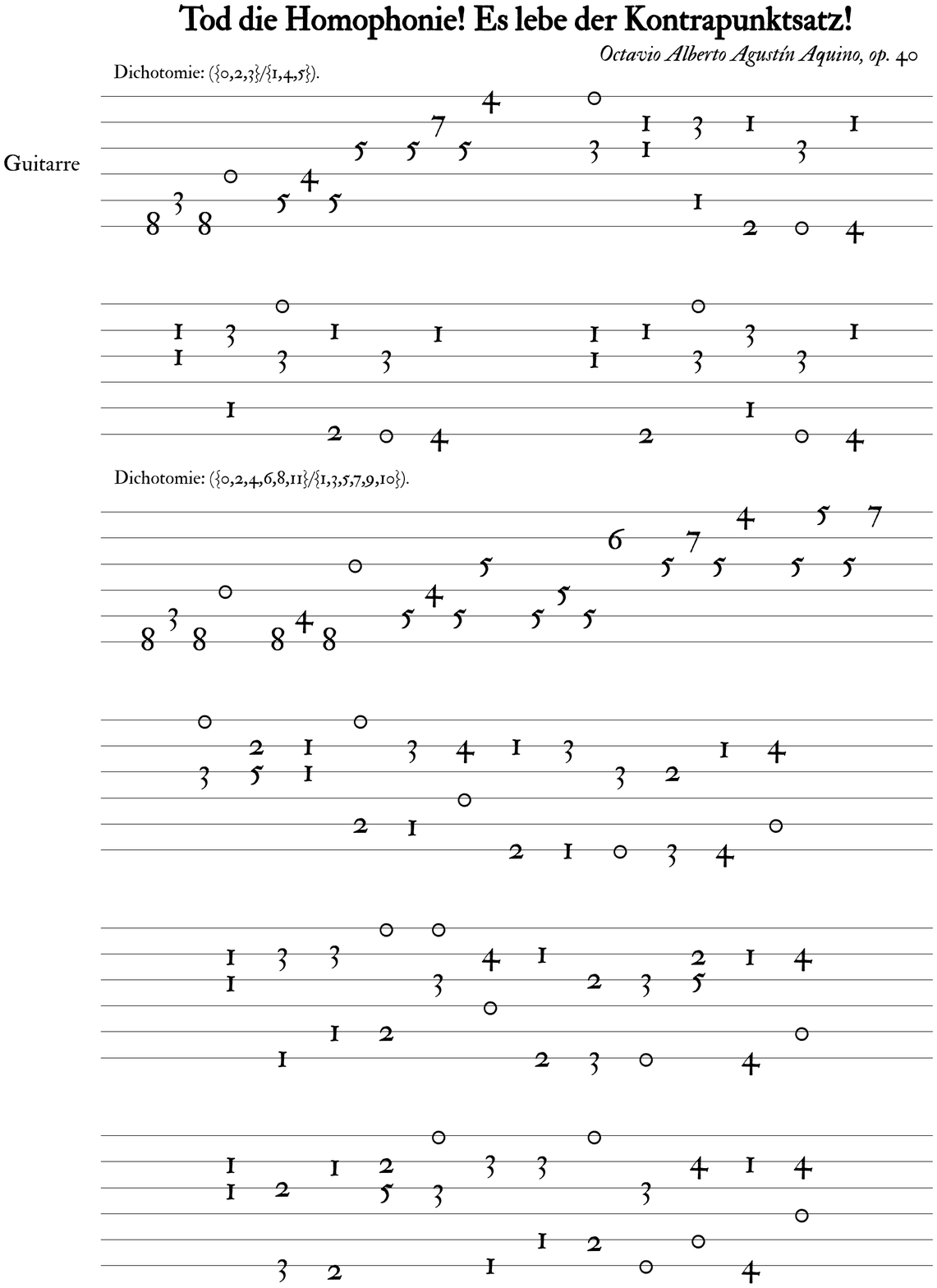}
}
\parbox{2.8in}{
\includegraphics[page=2,scale=0.37]{Agustin_Aquino_Op40.pdf}
}
\caption{Tod der Homophonie! Es lebe der Kontrapunktsatz!, op. 40, for guitar. Copyright by Octavio Agust\'{i}n-Aquino}
\end{figure}

The piece {\em Tod der Homophonie! Es lebe der Kontrapunktsatz!}, op. 40 (Figure \ref{fig:kontrapunktsatz}), is written for guitar, and it is an illustration of Theorems  \ref{T:KKS} and \ref{T:Embedding}, which appear in \cite{cpworlds}. It consists of three parts, one for each of the dichotomies $(\{0,2,3\}/\{1,4,5\}$, $(\{0,1,4,5,6,9\}/\{2,3,7,8,10,11\}$ (this is within the orbit of the $(K/D)$ dichotomy) and
$$
 ({0,1,2,3,5,8,9,10,11,12,15,18} , {4,6,7,13,14,16,17,19,20,21,22,23}).
$$ 

In each part the dichotomy is presented first as a series such that the interval between two consecutive tones is the image of the preceding interval under the polarity. Then a counterpoint in each world is displayed; the second part the $12$-tone missing consonances from the preceding world are shuffled between the original ones. Because of time constraints and difficulty of computation this scheme was not continued in the third part. A slide is required in order to play the $24$-tone part. This work was premiered October 8th, 2015 by the author \cite{youtubekontrapunktsatz}, during the presentation of the book. The score is a tablature.

\begin{figure}[h!]\label{fig:vivir}
\parbox{2.8in}{
\includegraphics[page=1,scale=0.3]{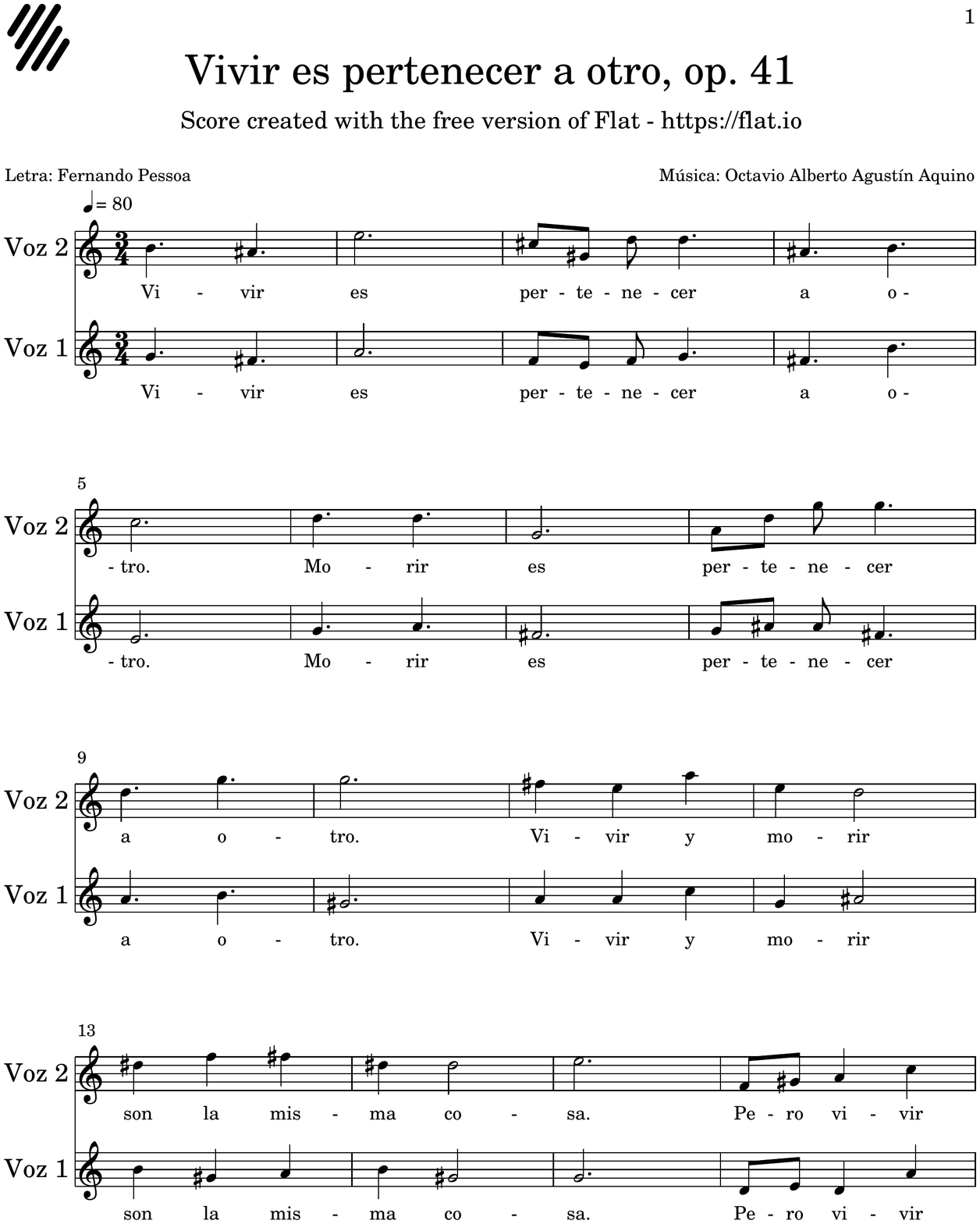}
}
\parbox{2.8in}{
\includegraphics[page=2,scale=0.3]{vivir.pdf}
}
\caption{Vivir es pertenecer a otro, op. 41, for two voices. Copyright by Octavio Agust\'{i}n-Aquino}
\end{figure}

{\em Vivir es pertenecer a otro}, op. 41 (Figure \ref{fig:vivir}), is a small polyphonic composition for two voices dedicated to the memory of Pierre Boulez, who died in 2016. The lyrics is a translation of a poem by Fernando Pessoa: ``Vivir es pertenecer a otro. Morir es pertenecer a otro. Vivir y morir son la misma cosa. Pero vivir es pertenecer a otro de afuera y morir es pertenecer a otro de adentro''. Two worlds of counterpoint are used. One is the classical Fuxian world and the other is the Ionian world. ``Life'' occurs within the Fuxian world, and ``death'' within the Ionian world. The ``belonging'' across worlds is represented by morphisms between them, as suggested by the text. Thus, life as ``belonging to someone outside'' is represented by a morphism from $(K/D)$ to $(I/J)$, whereas death as ``belonging to someone inside'' is represented by an automorphism from $(K/D)$ to $(K/D)$. In the middle part, when ``living'' and ``dying'' are claimed to be equal, the counterpoint corresponds to a succession of intervals valid in both worlds. The morphisms and checking of counterpoint steps were done with the BollyFux rubettes as well.

\begin{figure}[h!]\label{fig:morceau}
\parbox{2.8in}{
\includegraphics[page=1,scale=0.37]{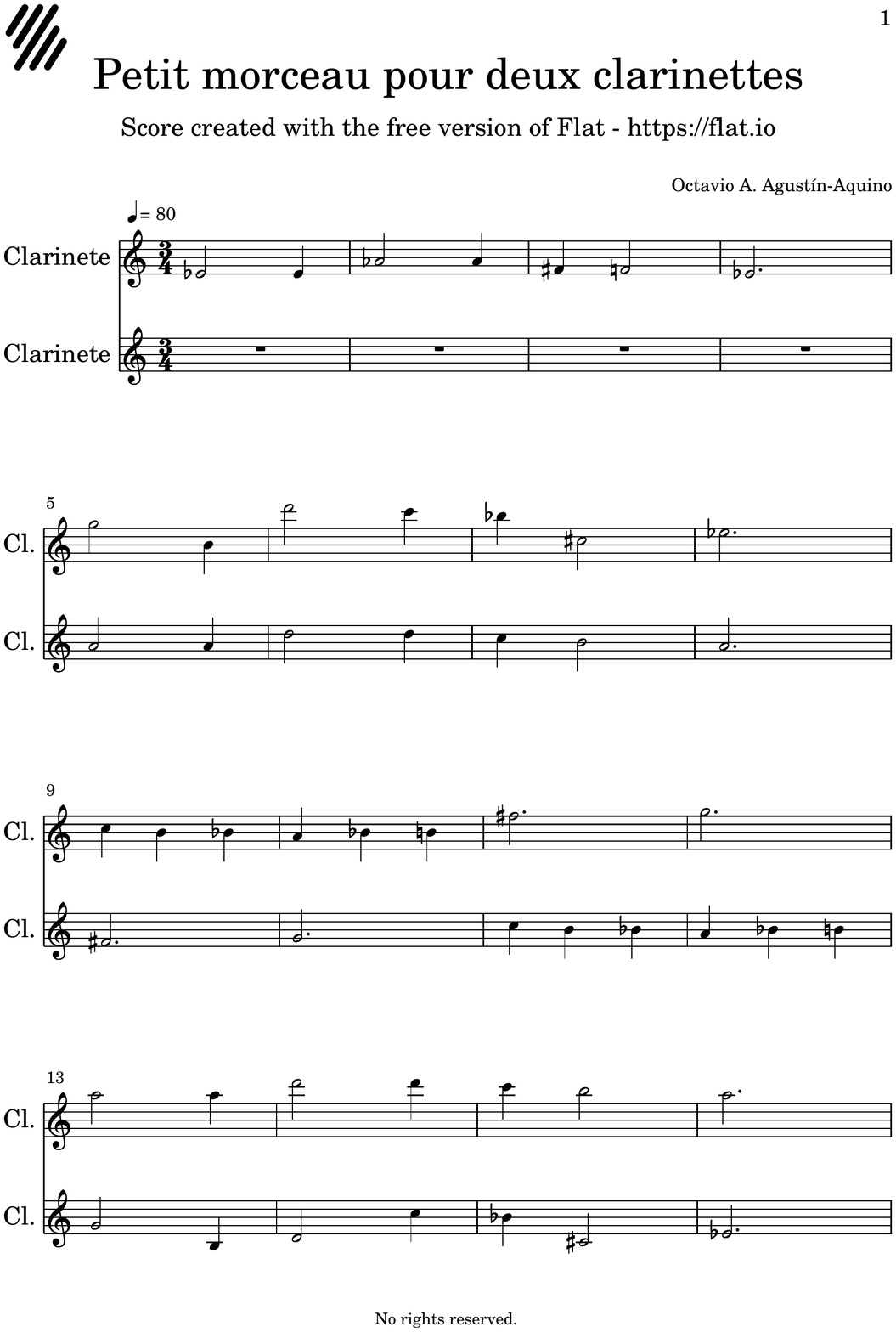}
}
\parbox{2.8in}{
\includegraphics[page=2,scale=0.37]{petit_morceau.pdf}
}
\caption{Petite morceau pour deux clarinettes, op. 42, for two clarinets. Copyright by Octavio Agust\'{i}n-Aquino}
\end{figure}

The {\em Petit morceau pour deux clarinettes}, op. 42 (Figure \ref{fig:morceau}), is a small fugue for two voices written for two clarinets and it is dedicated to the Museum of Louvre. It is written in Scriabin's mystic world dichotomy and it was chosen, in particular, because it allows for the double counterpoint required for the fugal treatment. The subject is taken from the medieval French song ``L'homme arm\'{e}''. It has an exposition, a counterexposition, two developments after each of the later and a closing stretto, following in part the model exposed in \cite[pp. 83-84]{mann}. It was composed semiautomatically with an Octave script that checks transitions as described in \cite{madridpaper}.

\begin{figure}[h!]\label{fig:archimedes}
\begin{center}
\includegraphics[page=1,scale=0.6]{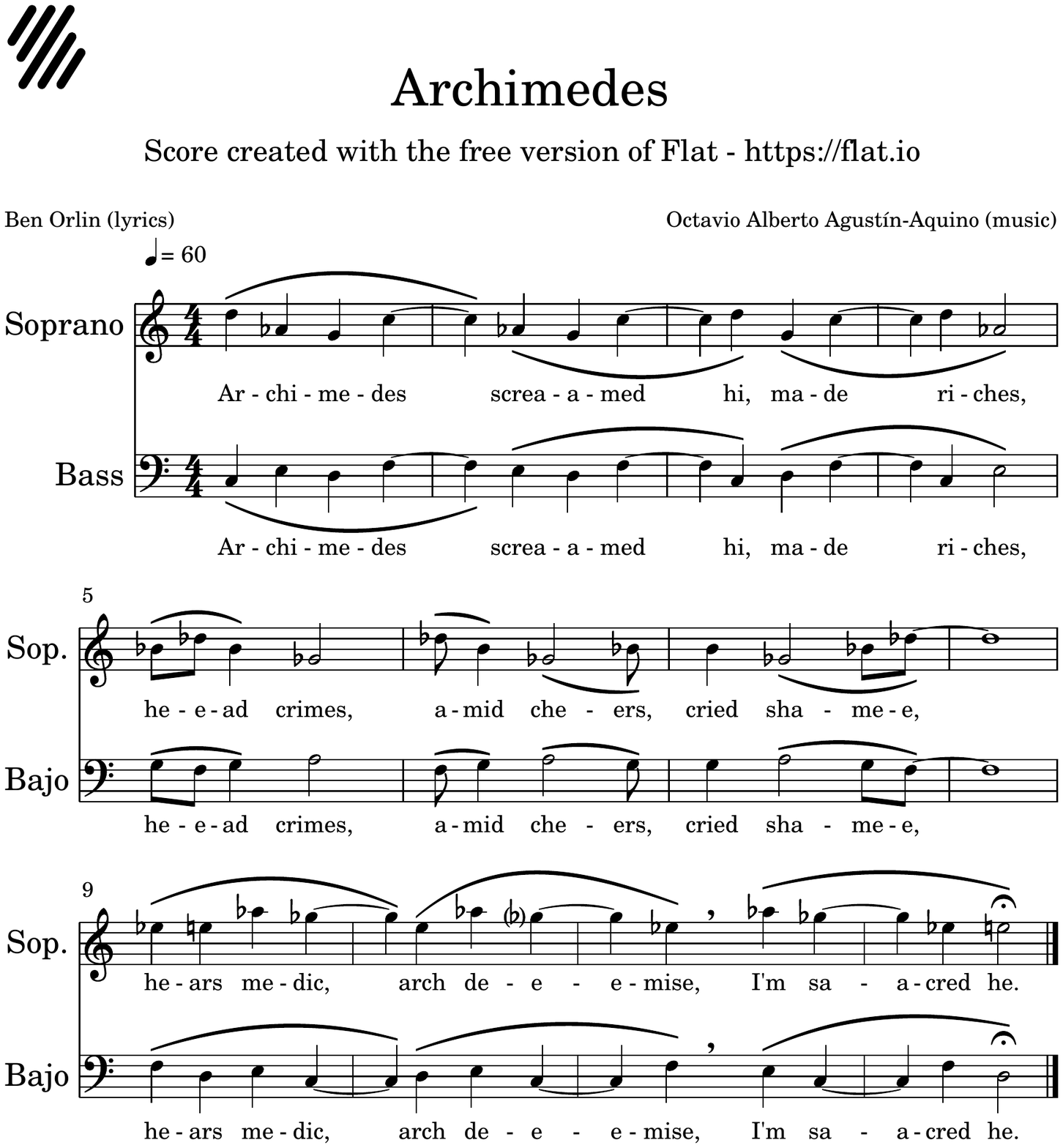}
\end{center}
\caption{Archimedes, op. 45, for two voices. Copyright by Octavio Agust\'{i}n-Aquino}
\end{figure}

The most recent composition, {\em Archimedes}, op. 45 (Figure \ref{fig:archimedes}), is a cantata with a libretto by Ben Orlin, which consists in eight anagrams of the name ``Archimedes'': ``Screamed hi. Made riches. Head crimes. Amid cheers. Cried shame. Hears medic. Arch demise. I'm sacred he''. Since we have nine anagrams, we divided them among three counterpoint worlds: Fuxian, Ionian and Mystic, in that order. The sequences of counterpoint intervals was composed such that the three for each world are permutations of the first, and all allowed by Mazzola's model. It was not composed using the BollyFux rubettes but the Octave script for op. 42, and thus the counterpoint in the different worlds are not morphed versions of one of them.

\subsection{Example by Jeffery Liu.}\label{EXPL_JEFF}
\begin{figure}[h!]\label{fig:sarabanda}
\parbox{2.8in}{
\includegraphics[page=10,scale=0.37]{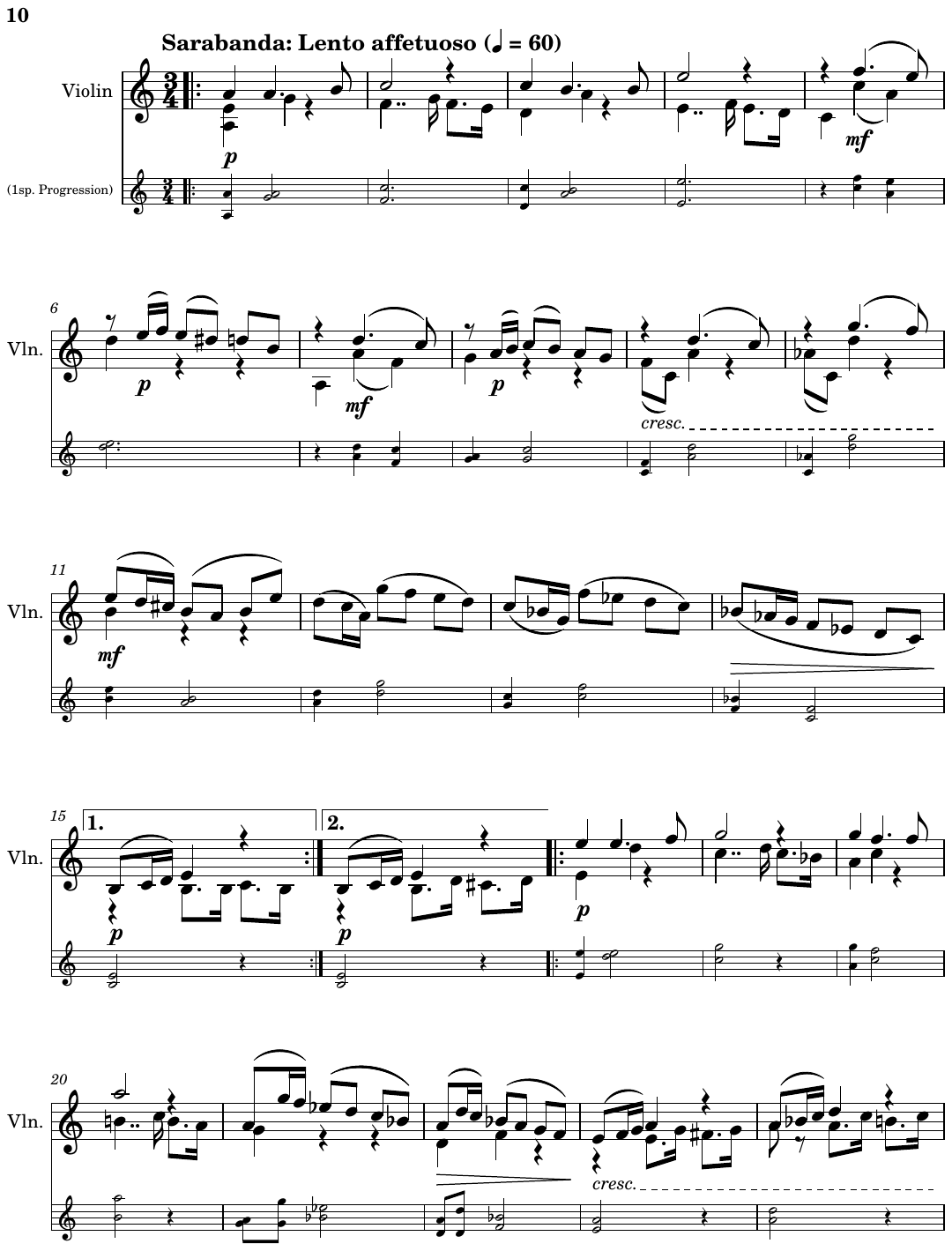}
}
\parbox{2.8in}{
\includegraphics[page=11,scale=0.37]{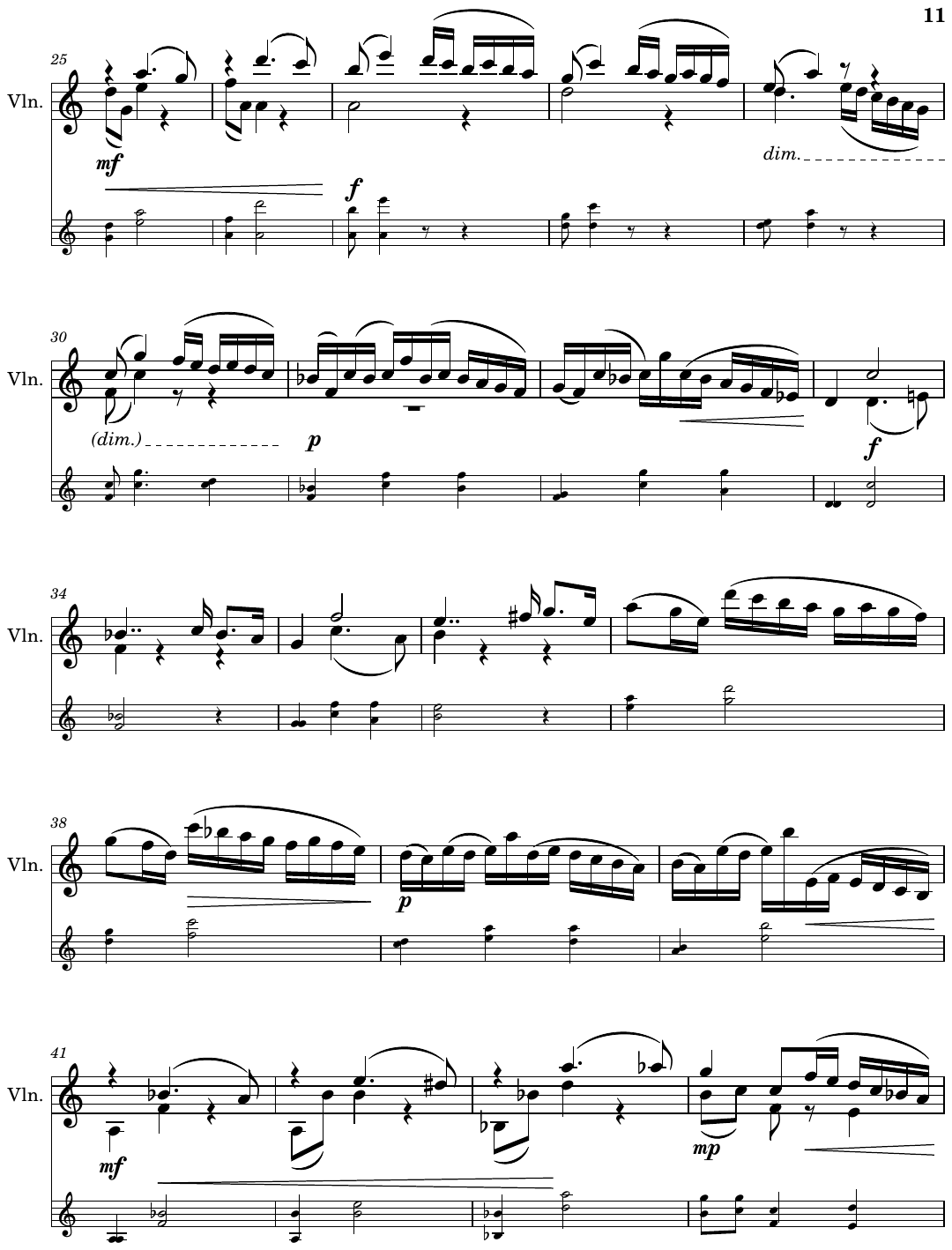}
}
\caption{Excerpt from Sarabanda of {\em Partita in A minor in the Counterpoint Dichotomy $\{0, 2, 5, 7, 8, 10\}$}. Copyright by Jeffery Liu.}
\end{figure}

The {\em Partita in A minor in the Counterpoint Dichotomy $\{0, 2, 5, 7, 8, 10\}$} for solo violin was newly composed as an application of the counterpoint model in a more elaborate style.\footnote{An online score and playback for the Partita in A minor can be found at\\ https://musescore.com/user/2028261/scores/6769737.\\ The PDF score and the mp3 file are also available for download at\\ https://drive.google.com/drive/folders/1y2HDBXS6KfNT6y9vECN1upfoKD4le3Zw?usp=sharing} The dichotomy $\{0, 2, 5, 7, 8, 10\}$ is a member of dichotomy class 64 by the numbering in Table \ref{tab:dichotomyClasses}. It was chosen for its compatibility with changing counterpoint orientation (when two voices exchange roles between sweeping and hanging counterpoint); it contains as subsets the three orbits $\{0\}$, $\{2, 10\}$, and $\{5, 7\}$ of $\ZZ_{12}$ under the group action of inversion, $T^0.11$, so that a majority of the consonances are preserved when the cantus firmus switches from the bottom to the top voice and thereby the intervals become inverted. In this regard, compare mm.\ 1-2 and mm.\ 32-33 in Figure \ref{fig:sarabanda}.  The underlying first species progression (given here in the small notes underneath the staff) was verified with the BollyWorld rubettes, but free elaborations such as passing tones, neighbor tones, and arpeggiation were permitted. Additionally, the melodic intervals also emphasize the consonant intervals of the dichotomy. A two-voice texture is implied on the solo violin by the use of double stops or by contrast in dynamics and register. The {\em Partita in A minor} has five movements: Preludio, Allemanda, Corrente, Sarabanda, Giga; these are the four standard dances of the Baroque suite preceded by a prelude. This choice of a standardized genre `controls' for the non-contrapuntal aspects of rhythm, meter, and form, since the four Baroque dances each have their own characteristic rhythms and meters, and are all in binary form. This isolates the dichotomy of intervals as the explanatory variable of its \textit{Affekt} and readily allows comparison with historical examples in classical counterpoint, such as the Partitas for Solo Violin (BWV 1002, 1004, 1006) by J.\ S.\ Bach.

\section{Proposal of a Poll/Inquiry About the Aesthetic Value of Such Compositions}\label{POLL}

The idea of this inquiry is to learn from the reader's impressions about their aesthetic judgement regarding these compositions in exotic counterpoint worlds. We shall propose the following list of questions which you might answer. The authors would be happy to receive your answers by email to the authors and evaluate your judgements and put them in a subsequent paper which the authors want to submit to this journal.
\begin{enumerate}
\item Did you appreciate these compositions in comparison to classical European compositions?
\item Do you believe the instrumentation is important for the aesthetic judgement?
\item Did you feel uncomfortable with some classically dissonant intervals?
\item Did you like or dislike the progression of new consonant intervals?
\end{enumerate}

\newpage
%




\bibliographystyle{plain}
\bibliography{tMAMguide}










\end{document}